\documentclass[12pt]{iopart}

\usepackage{graphicx}
\usepackage{wrapfig}
\usepackage{amssymb}

\begin{document}

\title[Quantum realization of extensive games]{Quantum realization of extensive
games}
\author{Piotr Fr\c{a}ckiewicz}
\address{Institute of Mathematics of the
Polish Academy of Sciences\\ 00-956, Warsaw, Poland}
\newtheorem{lemma}{Lemma}[section]
\newtheorem{definition}{Definition}[section]
\newtheorem{theorem}{Theorem}[section]
\newtheorem{example}{Example}[section]
\ead{P.Frackiewicz@impan.gov.pl}
\begin{abstract}
We generalize a concept of classical finite extensive game to make
it useful for application of quantum objects. The generalization
extends a quantum realization scheme of static games to any finite
extensive game. It represents an extension of any classical finite
extensive games to the quantum domain. In addition our model is
compatible with well-known quantum schemes of static games. The
paper is summed up by two examples.
\end{abstract}
\section{Introduction}
The quantum game theory is based on the combination of game theory
and quantum information. From the mathematical point of view an
arena of a~quantum game is a~tensor product of multidimensional
Hilbert spaces where unitary operators acting on fixed vectors
from these spaces are treated as actions taken by players
\cite{eisert2}. The paper \cite{eisert} deserves the special
attention. In this paper the authors introduce the model of
a~quantum realization of any static 2$\times$2 game. Another
important paper is \cite{marinatto} where the authors show an
alternative model of a quantum static game. Quantum game theory
goes beyond static games. Among recent papers there can be found
problems of duopoly, poker games, repeated games, etc., played in
the quantum domain. Motivation for our research has been a
scientific niche that remains in an area of quantum extensive
games. Our paper is entirely dedicated to this topic entirely.

We decided to remind in next section basic notions of game theory
which will be necessary in the sequel. Readers who are not
familiar with game theory are encouraged to consult \cite{myerson}
and \cite{osborne}. Necessary elements of quantum information can
be found in \cite{hirvensalo} and \cite{nielsen}.
\section{Preliminaries of game theory}
This section starts with defining a game in an extensive form. In
all cases when we refer to the concept of the game, we have in
mind finite games that is, games with finite set of strategies for
each player.
\begin{definition}{\rm \cite{osborne}} Let the following components be
given:
\begin{enumerate}
\item A finite set $N = \{1,2,\dots,n\}$ of players. \item A set
$H$ of finite sequences that satisfies the following two
properties:
\begin{enumerate}
\item The empty sequence $\emptyset$ is a member of $H$. \item If
$(a_k)_{k = 1,2,\dots, K} \in H$ and $L<K$ then $(a_k)_{k =
1,2,\dots, L} \in H$.
\end{enumerate}
Each member of $H$ is a history and each component of a history is
an action taken by a player. A history $(a_{1}, a_{2},\dots,
a_{K}) \in H$ is terminal if there is no $a_{K+1}$ such that
$(a_{1}, a_{2},\dots, a_{K}, a_{K+1}) \in H$. The set of actions
available after the nonterminal history $h$ is denoted $A(h) = \{a
\colon (h,a) \in H\}$ and the set of terminal histories is denoted
$T$. \item The player function $P \colon H \setminus  T
\rightarrow N \cup c$ that points to a player who takes an action
after the history $h$. If $P(h) = c$ then chance (the
chance-mover) determines the action taken after the history $h$.
\item A function $f_{c}$ that associates with each history $h$ for
which $P(h) = c$ an~independent probability distribution
$f_{c}(\cdot|h)$ on $A(h)$. \item For each player $i\in N$ a
partition $\mathcal{I}_{i}$ of $\{h \in H \setminus T: P(h) = i\}$
with the property that for each $I_{i} \in \mathcal{I}_{i}$ and
for each $h$, $h'$ $\in I_{i}$  an equality $A(h) = A(h')$ is
fulfilled.
\end{enumerate}
The extensive game form is a tuple $(N, H, P, f_{c},
\{\mathcal{I}_{i}\}_{i \in N})$. \label{-1}
\end{definition}
\noindent Each element $I_{i}$ of the partition $\mathcal{I}_{i}$
is called an information set for a player $i$. The partition of
set $H \setminus T$ into the information sets corresponds to the
state of players' knowledge. A player who makes move after certain
history $h$ belonging to an information set from
$\mathcal{I}_{i}$, knows that the current course of the game takes
the form of one of histories being part of $I_{i} \in
\mathcal{I}_{i}$. She does not know, however, if it is the history
$h$ or the other history from $I_{i}$.

The complete description of a game also requires a utility
function for each player~$i$ to be defined. The full description
of this function can be found in \cite{myerson} and
\cite{osborne}. For our purpose it suffices to understand the
utility function as a function assigning a~real number to each
terminal history. This number reflects preference of the $i$-th
player with respect to particular terminal history.
\begin{definition}{\rm \cite{osborne}}
The extensive game form together with a collection $\{u_{i}\colon
i\in N\}$ of~utility functions $u_{i} \colon T \to \mathbb{R}$ is
called the extensive game.\label{egame}
\end{definition} Although an extensive game from definition \ref{egame} describes
every extensive game, our deliberations focus on games with
perfect recall - this means games in which at each stage every
player remembers all the information that she knew earlier and all
of her own past moves. Further in the article if there is need to
depict utility values (payoffs) for all players, we shall consider
the utility function as function $u:T \to R^n$ defined as $u(h) =
(u_{i}(h))_{i=1,2,\dots,n}$.

A game as a mathematical model can be described in algebraic
language, thereby including a concept of isomorphism. Static game
is defined only by a set of players $N$, collection of sets of
strategies $\{S_{i}\}_{i \in N}$ and a utility function $u\colon
\prod_{i=1}^nS_{i} \to R^n$. For this reason games: $(N,
\{S_{i}\}_{i \in N}, u)$  and $(N, \{S'_{i}\}_{i \in N}, u')$ are
isomorphic iff there exists bijection $\zeta\colon
\prod_{i=1}^nS_{i} \to \prod_{i=1}^nS'_{i}$ such that $u(s) =
u'(\zeta(s))$. The players who are taking part in some game cannot
notice that they are playing indeed some of its isomorphic
equivalents. Contrary to static games, the notion of isomorphism
of extensive games is more complex.
\begin{definition}
The games in the form of tuples $( N, H, P, f_{c},
\{\mathcal{I}_{i}\}_{i \in N}, \{u_{i}\}_{i\in N})$ and $( N', H',
P', f'_{c}, \{\mathcal{I}'_{i}\}_{i \in N}, \{u'_{i}\}_{i\in N})$
are called isomorphic, if there exists a bijective function $\xi
\colon H \to~H'$ that satisfies following conditions:
\begin{enumerate}
\item $\xi(\emptyset) = \emptyset'$ \item $\forall_{h \in
H\setminus T}$ $\bigl(\xi(h,a) = (h',a') \Rightarrow \xi(h) =
h'\bigr)$, \item $\forall_{h \in H \setminus T}$ $P(h) =
P'(\xi(h))$ \item $\forall_{h \colon P(h) = c}$ $f_{c}(\cdot|h) =
f'_{c}(\cdot|\xi(h))$, \item $\forall_{i \in N} \forall_{I_{i} \in
\mathcal{I}_{i}}$ $\xi(I_{i}) = I'_{i}$, \item $\forall_{h \in T}
\forall_{i \in N}$ $u_{i}(h) = u'_{i}(\xi(h))$.
\end{enumerate} \label{isomorphism}
\end{definition}
\noindent We realize that the above defined notion of isomorphic
games contains very restrictive conditions. According to the
definition mentioned above, isomorphic games differ only by marks
of all their components. The case of two games in which one of
them will be modified by addition only one action $a_{1}$ of some
player to the beginning history $\emptyset$ so that $A(\emptyset)
= \{a_{1}\}$, does not satisfy the conditions of Definition
\ref{isomorphism}.

The notions: action and strategy mean the same in static games,
because the players choose their actions once and simultaneously.
In the majority of extensive games players can make their decision
about an action depending on all the actions taken previously by
themselves and also by all the other players. In other words,
players can have some plans of actions at their disposal such that
these plans point out to a specific action depending on the course
of a game. Such a plan is defined as a (pure) strategy in an
extensive game.
\begin{definition}{\rm \cite{osborne}}
A strategy $s_{i}$ of player $i$ in a game $( N, H, P, f_{c},
\{\mathcal{I}_{i}\}_{i \in N}, u)$ is a~function that assigns an
action in $A(I_{i})$ to each information set $I_{i} \in
\mathcal{I}$. \label{strategy}
\end{definition}
A sequence of strategies of all players $s=(s_{1},
s_{2},\dots,s_{n})$ is called a profile of the strategies. Each
profile $s$ determines unambiguously (if the chance mover has been
excluded) some terminal history and each of its subhistories.
However, for a fixed history $h$ there may exist a lot of
strategies' profiles which generate $h$. Generally:
\begin{definition}{\rm \cite{osborne}}
A strategy $s_{i}$ of a player $i$ is consistent with some history
$h = (a_{1}, a_{2},\dots,a_{k})$ if for every (strict) subhistory
$(a_{1}, a_{2},\dots,a_{l})$ of $h$ for which $P(a_{1},
a_{2},\dots,a_{l}) = i$ the condition $s_{i}(a_{1},
a_{2},\dots,a_{l}) = a_{l+1}$ is fulfilled. We define a~profile $s
= (s_{i})_{i =1,2,\dots,n}$ to be consistent with $h$ iff for
every player $i$, $s_{i}$ is consistent with~$h$. \label{zgodnosc}
\end{definition}
Definitions \ref{strategy} and \ref{zgodnosc} imply that any
extensive game defined by Definition \ref{egame} induces some
static (strategic) game $(N, \{S_{i}\}_{i \in N}, u)$. Then for
each $i \in N$ a set $S_{i}$ is the set of all possible functions
defined by Definition \ref{strategy}. The utility function $u$ is
the same as the one in the extensive game but is redefined from a
domain $T$ to the domain $\prod_{i=1}^nS_{i}$ using the notion
defined in Definition \ref{zgodnosc}. It follows that notions
assigned to static games are used in extensive games. One of these
most important notions is a notion of an equilibrium introduced by
John Nash in \cite{nash}:
\begin{definition}
Let mark by $( N, S_{i}, \{u_{i}\}_{i \in N} )$ a strategic form
of an extensive game with perfect recall. A profile of strategies
$(s^*_{1}, s^*_{2},\dots,s^*_{n})$ is a Nash equilibrium if for
each player $i \in N$:
\begin{equation}
u_{i}(s^*_{1},s^*_{2},\dots,s^*_{n}) \geqslant
u_{i}(s^*_{1},s^*_{2},\dots,s^*_{i-1},s_{i},s^*_{i+1},\dots,s^*_{n})
~~ \mbox{for all} ~~ s_{i} \in S_{i}. \label{nashequation}
\end{equation}
\end{definition} Inequality (\ref{nashequation}) means that Nash equilibrium  is a
profile where the strategy of each player is optimal if we accept
the choice of its opponents to be fixed. In the equilibrium none
of the players has any reason to unilaterally deviate from an
equilibrium strategy.
\section{Eisert's generalized scheme for two-person static quantum game}
The quantum model described in this paper comes originally from
\cite{eisert}, but we will describe its general version based on
\cite{nawaz}, with a little change of a space's base however.
Quantization of a two-person static game begins with preparation
of an initial state described by a vector from the space
$\mathbb{C}^2 \otimes \mathbb{C}^2$ with a standard base
$\{|00\rangle, |01\rangle, |10\rangle, |11\rangle\}$:
\begin{equation}
|\Psi\rangle = \cos\frac{\gamma}{2}|00\rangle +
i\sin\frac{\gamma}{2}|11\rangle \quad \rm{for} \quad \gamma \in
[0, \pi]. \label{initialstate}
\end{equation}
Each of the players has at his disposal a~set of unitary operators
depending on two parameters $\theta$ and $\phi$ of the form:
\begin{equation}
U_{j} = \cos\frac{\theta_j}{2}J_{j} + \sin\frac{\theta_j}{2} C_{j}
\quad \mbox{for} \quad \theta_{j} \in [0, \pi], \label{operator}
\end{equation}
where $J_{j}$ and $C_{j}$ are defined as follows:
\begin{eqnarray}
J_{j}|0\rangle &= e^{i\phi_j}|0\rangle, \quad J_{j}|1\rangle& =
e^{-i\phi_j}|1\rangle \quad
\mbox{for} \quad \phi_j \in [0, \frac{\pi}{2}]; \nonumber\\
C_j|0\rangle &= -|1\rangle, \quad ~C_j|1\rangle& = |0\rangle
\qquad ~~~ \mbox{and} \quad j =1,2. \label{operator2}
\end{eqnarray}
Operators (\ref{operator}) are treated as actions taken in the
game where players act, respectively, on the first and the second
qubit in the initial state (\ref{initialstate}):
\begin{equation}
|\Psi_{fin}\rangle = (U_{1} \otimes U_{2})|\Psi\rangle.
\label{state}
\end{equation}
In accordance with formulas (\ref{initialstate}) to
(\ref{operator2}), the state (\ref{state}) takes form:
\begin{equation}
|\Psi_{fin}\rangle = \chi_{00}|00\rangle + \chi_{01}|01\rangle +
\chi_{10}|10\rangle + \chi_{11}|11\rangle, \label{state2}
\end{equation}
where elements $\chi_{kl}$ are specified by equalities:
\begin{eqnarray}
\chi_{00} &=& e^{i(\phi_{1} +
\phi_{2})}\cos\frac{\gamma}{2}\cos\frac{\theta_{1}}{2}\cos\frac{\theta_{2}}{2}
+
i\sin\frac{\gamma}{2}\sin\frac{\theta_{1}}{2}\sin\frac{\theta_{2}}{2};
\nonumber
\\
\chi_{01} &=&
-e^{i\phi_{1}}\cos\frac{\gamma}{2}\cos\frac{\theta_{1}}{2}\sin\frac{\theta_{2}}{2}
+
ie^{-i\phi_{2}}\sin\frac{\gamma}{2}\sin\frac{\theta_{1}}{2}\cos\frac{\theta_{2}}{2}; \nonumber\\
\chi_{10} &=&
-e^{i\phi_{2}}\cos\frac{\gamma}{2}\sin\frac{\theta_{1}}{2}\cos\frac{\theta_{2}}{2}
+
ie^{-i\phi_{1}}\sin\frac{\gamma}{2}\cos\frac{\theta_{1}}{2}\sin\frac{\theta_{2}}{2};
\nonumber
\\ \chi_{11} &=&
\cos\frac{\gamma}{2}\sin\frac{\theta_{1}}{2}\sin\frac{\theta_{2}}{2}
+ ie^{-i(\phi_{1} +
\phi_{2})}\sin\frac{\gamma}{2}\cos\frac{\theta_{1}}{2}\cos\frac{\theta_{2}}{2}.
\label{rozpiska2}
\end{eqnarray}
The utility function for the players is defined by operator:
\begin{equation}
X = \Delta_{00}|00\rangle\langle00| +
\Delta_{01}|01\rangle\langle01| + \Delta_{10}|10\rangle\langle10|
+ \Delta_{11}|11\rangle\langle11|, \label{rozpiska}
\end{equation} where each element $\Delta_{kl}$ is a
two-dimensional payoff vector with real value entries defined by
an `original' game. If a quantum state (\ref{state}) is presented
in the form of density matrix $\rho_{fin} =
|\Psi_{fin}\rangle\langle\Psi_{fin}|$ payoffs that players gain
are expressed by the following formula:
\begin{equation}
\pi\bigl((\theta_{1}, \phi_{1}), (\theta_{2}, \phi_{2})\bigr) =
\mbox{Tr}(X \rho_{fin}), \label{payoff}
\end{equation}
which by using equalities (\ref{state2}) to (\ref{rozpiska}),
 can be expressed as:
\begin{equation}
\pi\bigl((\theta_{1}, \phi_{1}), (\theta_{2}, \phi_{2})\bigr) =
\sum_{k,l \in \{0,1\}} \Delta_{kl}|\chi_{kl}|^2. \label{payoff2}
\end{equation}
\section{Quantum realization of extensive game}
In this section we define an extensive game played with the use of
quantum objects (qudits).  To simplify the idea we restricted our
consideration to games without the chance-mover (typically called
Nature). However in further part of the paper we will point out
how to extend our scheme to games with action taken by Nature.

Let us assume that we have $m$ various quantum system, each
described with the use of a space $\mathbb{C}^{d_{j}}$ spanned by
the orthonormal base $\mathcal{B}^{d_{j}} = \{|0\rangle,
|1\rangle,\dots,|d_{j}-1\rangle\}$ for $j = 1,2,\dots,m$.
Furthermore, let's consider an initial quantum state of a
composite system described by a unit vector $|\Psi\rangle$ from
the tensor product $\mathbb{C}^{d_{1}} \otimes \mathbb{C}^{d_{2}}
\otimes \dots \otimes \mathbb{C}^{d_{m}}$ with the base
$\bigl\{|\nu_{1}\rangle \otimes |\nu_{2}\rangle \otimes \dots
\otimes |\nu_{m}\rangle\ \colon |\nu_{j}\rangle \in
\mathcal{B}^{d_{j}}\}$. The vector $|\Psi\rangle$ takes the form:
\begin{equation}
|\Psi\rangle =
\sum_{\nu_{1}=0}^{d_{1}-1}\sum_{\nu_{2}=0}^{d_{2}-1}\dots\sum_{\nu_{m}=0}^{d_{m}-1}\alpha_{\nu_{1},\nu_{2},\dots,\nu_{m}}|\nu_{1},\nu_{2},\dots,\nu_{m}\rangle
\label{0}.
\end{equation}
Additionally, let's mark by:
\begin{enumerate}
\item[{\em(i)}] $\bigl\{\mathcal{U}_{1},
\mathcal{U}_{2},\dots,\mathcal{U}_{m}\}$ a collection of sets of
unitary operators. Each set $\mathcal{U}_{j}$ is a subset of the
set $\mathsf{SU}(d_{j})$ including $d_{j}$ operators $V_{0},
V_{1},\dots, V_{d_{j}-1}$ defined as follows:
\begin{equation} V_{t}|\nu_{j}\rangle =
e^{i\phi_{t}}|\nu_{j} \oplus t\rangle ~~ \mbox{for any}~~ t \in
\{0,1,\dots,d_{j}-1\},
 \label{operatorybazowe} \end{equation} where $|\nu_{j}\rangle \in \mathcal{B}^{d_{j}}$, $e^{i\phi_{t}}$ is some phase factor and $\oplus$ means addition
modulo $d_{j}$. \item[{\em(ii)}] $\{M_{\nu_{j}}\}$ a collection of
operators that provides the description of quantum measurements.
An operator $M_{\nu_{j}}$ for $j = 1,2,\dots,m$ and $\nu_{j} =
0,1,\dots,d_{j}-1$ is expressed by the formula:
\begin{equation}
M_{\nu_{j}} = I^{\otimes j-1} \otimes |\nu_{j}\rangle
\langle\nu_{j}| \otimes I^{\otimes m-j}. \label{pomiar2}
\end{equation}
\end{enumerate}
The measurement operator $M_{\nu_{j}}$ defines measurement outcome
$\nu_{j}$ on a single qudit~$j$ composing (\ref{0}).

Let us assume abbreviated notation $\{U_{j}, \nu_{j}\}$ for a
unitary operation $U_j \in \mathcal{U}_{j}$ carried out on $j$-th
qudit:
\begin{equation}
I^{\otimes j-1} \otimes U_{j} \otimes I^{\otimes m-j}|\Psi\rangle
= |\Psi_{U_{j}}\rangle, \label{IUI}
\end{equation}
and a measurement outcome $\nu_{j}$ obtained after measurement on
this qudit. In accordance with von Neumann measurements the
post-measurement state $|\Psi_{\{U_{j}, \nu_{j}\}}\rangle$ is
\begin{equation}
\frac{M_{\nu_{j}}|\Psi_{U_{j}}\rangle}{\sqrt{\langle \Psi_{U_{j}}|
M_{\nu_{j}}|\Psi_{U_{j}}\rangle}} ~~ \mbox{with probability} ~~
\Pr(\nu_{j}) = \langle \Psi_{U_{j}}|
M_{\nu_{j}}|\Psi_{U_{j}}\rangle. \label{aftermeasurement}
\end{equation}
When there is no need to detail both of components from couple
$\{U_{j}, \nu_{j}\}$ we will write merely $q_{j} \doteq \{U_{j},
\nu_{j}\}$ to denote that a unitary operation $U_{j}$ and a
measurement yielding the result $\nu_{j}$ are performed on qudit
$j$. We define in a recurrent way a (finite) sequence $(q_{j_{1}},
q_{j_{2}},\dots,q_{j_{\lambda}})$ of operations on qudits
$j_{1},j_{2},\dots,j_{\lambda}$. Here, each element
$q_{j_{\kappa}}$ is a couple $\{U_{j_{\kappa}},
\nu_{j_{\kappa}}\}$ of operations of the initial state (\ref{0})
when the sequence of operations $(q_{j_{1}},
q_{j_{2}},\dots,q_{j_{\kappa -1}})$ on this state has occurred.

Given that a unitary operation $U_{j}$ has been chosen, a
probability $\Pr{(\{U_{j}, \nu_j\})}$ that couple $\{U_{j},
\nu_j\}$ occurs is independent of operation taken by the other
players on their own qudits. Therefore, following the recurrent
expression of a sequence $(q_{j_{\kappa}})_{\kappa =
1,2,\dots,\lambda}$ we define the probability to be
$\Pr((q_{j_{\kappa}})_{\kappa = 1,2,\dots,\lambda})) \equiv
\prod_{\kappa=1}^\lambda\Pr(\nu_{j_{\kappa}})$.
\subsection{Quantum
extensive game form} Let mark by $\ell_{\leqslant m}$ the set of
all possible subsequences $(q_{j_{\kappa}})_{\kappa =
1,2,\dots,\lambda}$ of sequences of the form $(q_{1},
q_{2},\dots,q_{m})$. Let define the collection in the power set
$\mathcal{P}(\ell_{\leqslant m})\colon$
\begin{equation}
\mathcal{L} = \biggl\{[(\nu_{j_{\kappa}})_{\kappa =
1,2,\dots,\lambda}] \colon
\begin{array}{l} (j_{\kappa})_{\kappa = 1,2,\dots,\lambda} ~
\mbox{is a subsequence of} ~ (j)_{j=1,2,\dots,m}\\
\nu_{j_{\kappa}} \in \{0,1,\dots,d_{j_{\kappa}}-1\}
\end{array}\bigg\},
\end{equation}
where each set of the form $[(\nu_{j_{\kappa}})_{\kappa =
1,2,\dots,\lambda}]$ consists of all sequences of operations
 $(q_{j_{\kappa}})_{\kappa =
1,2,\dots,\lambda}$ on which the sequence of measurement outcomes
$(\nu_{j_{\kappa}})_{\kappa = 1,2,\dots,\lambda}$ has occurred.
Notice that $\mathcal{L}$ is a~collection of pairwise disjoint
sets and equal to $\ell_{\leqslant m}$ in total. Moreover, every
set $[(\nu_{j_{\kappa}})]$ is nonempty (as unitary operations
include operations of the form (\ref{operatorybazowe})). It
entails that $\mathcal{L}$ is a partition of $\ell_{\leqslant m}$.
The partition determines unique equivalence relation for which
every set $[(\nu_{j_{\kappa}})]$ is an equivalence class. An
extensive game played on qudits is the game played according to
scenario from Definitions \ref{-1} and \ref{egame} except that it
is expressed in the language of classes from $\mathcal{L}$:
\begin{definition}
Quantum extensive game form consists of the following components:
\begin{enumerate}
\item A finite set of players $N=\{1,2,\dots,n\}$. \item An
initial state $|\Psi\rangle \in \mathbb{C}^{d_{1}} \otimes
\mathbb{C}^{d_{2}} \otimes \dots \otimes \mathbb{C}^{d_{m}}$ and
$m \geqslant n$. \item A collection $\{\mathcal{U}_{1},
\mathcal{U}_{2},\dots,\mathcal{U}_{m}\}$ of sets of unitary
operators. \item A subcollecton  $\mathcal{H} \subset \{\emptyset,
\mathcal{L}\}$ that fulfils the following three properties:
\begin{enumerate} \item $\emptyset \in \mathcal{H}$,
\item if $[(\nu_{j_{\kappa}})_{\kappa = 1,2,\dots,\lambda}] \in
\mathcal{H}$ then for any number $\iota \in \{1,2,\dots,
\lambda\}$ it implies that $\{[(\nu_{j_{\kappa}})_{\kappa =
1,2,\dots,\iota}] \colon \nu_{j_{\iota}} =
0,1,\dots,d_{j_{\iota}}-1\} \subset \mathcal{H}$ \item if~
$[(\nu_{j_{\kappa}})_{\kappa = 1,2,\dots,\lambda}] \in
\mathcal{H}$ and $j' \ne j_{\lambda}$ then $[(\nu_{j_{1}},
\nu_{j_{2}},\dots,\nu_{j_{\lambda-1}}, \nu_{j'})] \notin
\mathcal{H}$.
\end{enumerate}
Each sequence $(q_{j_{\kappa}})$ from a set $\bigcup(\mathcal{H})$
(i.e. a set of all sequences of $\mathcal{H}$) is called
a~history. The set $\emptyset$ is the initial history for which it
is assumed $\Pr(\emptyset) = 1$. A history $(q_{j_{\kappa}}) \in
[(\nu_{j_{\kappa}})_{j_{\kappa} = 1,2,\dots,j_{\lambda}}]$ (a
class $[(\nu_{j_{\kappa}})_{j_{\kappa} = 1,2,\dots,j_{\lambda}}]$)
is terminal if there is no number $j'$ such that $[(\nu_{j_{1}},
\nu_{j_{2}},\dots,\nu_{j_{\lambda}},\nu_{j'})] \in \mathcal{H}$.
The collection of all terminal classes will be
denoted~$\mathcal{T}$. A partition of a collection $\mathcal{H}$
into terminal and nonterminal classes defines a map
$\widetilde{A}$ on $\bigcup(\mathcal{H} \setminus \mathcal{T})$
such that $\widetilde{A}\left([(\nu_{j_{\kappa}})_{\kappa =
1,2,\dots,\lambda}]\right) = \{q_{j_{\lambda + 1}}\}$. A~set of
unitary operators $\mathcal{U}_{j_{\lambda + 1}}$ defined by a
couple $q_{j_{\lambda + 1}}$ is a set of available actions to
a~player whose turn is next after history from
$[(\nu_{j_{\kappa}})_{\kappa = 1,2,\dots,\lambda}]$. Each
measurement outcome $\nu_{j_{\lambda + 1}}$ corresponding to an
individual base state from $\mathcal{B}^{d_{j_{\lambda+1}}}$ is
treated as an outcome of some action $U_{j_{\lambda + 1}} \in
\mathcal{U}_{j_{\lambda + 1}}$. \item A player function
$\widetilde{P} \colon \bigcup(\mathcal{H} \setminus  \mathcal{T})
\to N$ with the property that a set
$\widetilde{P}([(\nu_{j_{\kappa}})])$ is a~singleton for every
class $[(\nu_{j_{\kappa}})] \in \mathcal{H} \setminus
\mathcal{T}$. It points at a player who takes an action when a
history from class $[(\nu_{j_{\kappa}})]$ has happened. A function
$\widetilde{P}$ together with a map $\widetilde{A}$ defines for
each player~$i$ a partition of collection
$\mathcal{H}\setminus\mathcal{T}$ into information sets
$\mathcal{J}_{i}^j$
 that take the form
$\{[(\nu_{j_{\kappa}})]~\in~\mathcal{H}\setminus\mathcal{T} \colon
\widetilde{P}([(\nu_{j_{\kappa}})]) =~\{i\} \wedge
\widetilde{A}([(\nu_{j_{\kappa}})]) = \{q_j\} \}$.
\end{enumerate} \label{qextensiveform}
\end{definition}
The quantum extensive game form $(N, |\Psi\rangle,
\{\mathcal{U}_j\}, \mathcal{H}, \widetilde{P})$ and the classical
extensive game form are almost identical with accuracy of
components' nature that they make up both the forms. The notion of
player $i$'s strategy in the extensive game also adapts for the
quantum case in natural way, i.e. it is a map that assigns a
unitary operator associated with a couple
$\widetilde{A}(\mathcal{J}_{i}^{j})$ to each information set. In
other words, a strategy of player $i$ determines exactly one
operation on each of qudits that are available for a player $i$.
There is a deep difference in a structure of a history though. In
normal extensive games actions taken by players unambiguously
determine some history. In the quantum case a~sequence (being a
history in quantum game) in which every element consists of
a~unitary operation on quantum object and then a measurement on
it, forms a history. It implies that on the one hand there are
more than one sequence of a unitary operation that correspond to
the same sequence of measurement outcomes. On the other hand as a
rule a given sequence of unitary operations and a measurement
performed on arbitrary qudits do not unambiguously sequences of
measurement outcomes. So now, to formulate anew the notion
presented in Definition \ref{zgodnosc}, we define a strategy
$s_{i}$ of player $i$ to be consistent with a history
$(q_{j_{1}},q_{j_2},\dots,q_{j_{\lambda}})$ if for every
subhistory $(q_{j_{1}}, q_{j_{2}},\dots,q_{j_{\kappa}})$ for which
$P(q_{j_{1}}, q_{j_{2}},\dots,q_{j_{\kappa}}) = i$ we have
$q_{j_{\kappa+1}} = \{s_{i}(q_{j_{1}},
q_{j_{2}},\dots,q_{j_{\kappa}}), \nu_{j_{\kappa+1}}\}$. The notion
formulated in this way implies that if a profile strategy $s$ is
given then there are more than one sequence $(q_{j_{\kappa}})$
that $s$ is consistent with $(q_{j_{\kappa}})$. The set of all
these sequences consists of all sequences in which unitary
operation are determined by the profile $s$.

In a tuple $(N, H, P, f_{c}, \{\mathcal{I}_{i}\}_{i \in N})$ which
describes in an explicit way some some extensive game, a partition
into information sets $\{\mathcal{I}_{i}\}_{i \in N}$ can be
omitted. It suffices to include in set $H$ description information
indicating which of nonterminal histories $h \in H$ implicate the
same set of action $A(h)$ (e.g. through applying appropriate
indices). The component $\mathcal{H}$ of a tuple $(N,
|\Psi\rangle, \{\mathcal{U}_j\}, \mathcal{H}, \widetilde{P})$
specifies all the histories which are predecessors of an operation
on the same qudit of initial state $|\Psi\rangle$, hence the lack
of elements $\mathcal{J}_i$ in the tuple. Notice more that
contrary to a classic extensive game the smallest information set
consists of one history. In the quantum extensive form the
smallest information set is composed of all the sequences in which
the same sequence of outcomes has occurred.

The scheme can be widened by the special player called Nature. For
instance, let us assume that after each of certain class of
histories a move of the Nature occurs and the number of her
possible alternatives is $d_{c}$. Then an initial state
characterizing so modified game is in the form of $|\Psi\rangle
\otimes |\psi\rangle \in \bigotimes_{j=1}^m \mathbb{C}^{d_{j}}
\otimes \mathbb{C}^{d_{c}}$. A measurement of qudit $|\psi\rangle$
generates independent probability distribution and a measurement
outcome of this qudit represents an `action' of Nature.
\subsection{Utility function}
The means of defining preferences of players in our scheme are
based on the well-known structure including \cite{eisert2} and
\cite{marinatto}.

Let $(N, |\Psi\rangle, \{\mathcal{U}_{j}\}, \mathcal{H},
\widetilde{P},)$ be a quantum extensive form and assign each class
$[(\nu_{j_{\kappa}})_{\kappa=1,2,\dots,\lambda}]$ from
$\mathcal{T}$ with a projector
\begin{equation} M_{[(\nu_{j_{\kappa}})]} = \sum_{\nu_{j} \colon j \notin \{j_{1}, j_{2},\dots,
j_{\lambda}\}}|\nu_{1},\nu_{2},\dots,\nu_{m}\rangle\langle
\nu_{1},\nu_{2},\dots,\nu_{m}|. \label{projector}
\end{equation}
Similarly to the scheme of quantum static games from the third
section we define a~payoff operator as
\begin{equation}
X_{e} = \sum_{[(\nu_{j_{\kappa}})] \in \mathcal{T}}
\Delta_{[(\nu_{j_{\kappa}})]}M_{[(\nu_{j_{\kappa}})]},
\label{epayoffoperator}
\end{equation}
where $\Delta_{[(\nu_{j_{\kappa}})]} = (\delta_{1},
\delta_{2},\dots,\delta_{n})$ is an element of $\mathbb{R}^{n}$
and each coordinate $\delta_{i}$ means a utility payoff for $i$-th
player. Let $\rho_{(q_{j_{\kappa}})}$ denote the density matrix of
the initial state $|\Psi\rangle$ given that a sequence
$(q_{j_{\kappa}}) \in \bigcup (\mathcal{T})$ on state
$|\Psi\rangle$ has occurred. Then $\widetilde{u}(q_{j_{\kappa}}) =
\mbox{Tr}(X_{e}\rho_{(q_{j_{\kappa}})})$. In the quantum extensive
game an expression $\mbox{Tr}(X_{e}\rho_{(q_{j_{\kappa}})})$ turns
out to be more simplified.

Suppose that $[(\nu_{\alpha_{\kappa}})_{\kappa =
1,2,\dots,\lambda}]$ and $[(\nu'_{\beta_{\kappa}})_{\kappa =
1,2,\dots,\lambda'}]$ are disjoint terminal classes
from~$\mathcal{T}$. Let $M_{[(\nu_{\alpha_{\kappa}})]}$ and
$M_{[(\nu'_{\beta_{\kappa}})]}$ be the projectors
(\ref{projector}) that correspond to these classes. Let us
estimate the product
$M_{[(\nu_{\alpha_{\kappa}})]}M_{[(\nu'_{\beta_{\kappa}})]}$. It
will be the zero operator if $M_{[(\nu_{\alpha_{\kappa}})]}$ and
$M_{[(\nu'_{\beta_{\kappa}})]}$ do not have any element $|\nu_{1},
\nu_{2},\dots,\nu_{m}\rangle\langle
\nu_{1},\nu_{2},\dots,\nu_{m}|$ in common. The condition $(c)$ of
Definition \ref{qextensiveform} implies that an equality
$\alpha_{1} = \beta_{1}$ must be true for any class
$[(\nu_{\alpha_{\kappa}})]$ and $[(\nu'_{\beta_{\kappa}})]$. Now,
if $\nu_{\alpha_{1}} \ne \nu'_{\alpha_{1}}$ then
$M_{[(\nu_{\alpha_{\kappa}})]}$ and
$M_{[(\nu'_{\beta_{\kappa}})]}$ vary in at least outcome of
$\alpha_{1}$ and it entails the product of 0. So, classes
$[(\nu_{\alpha_{1}},
\nu_{\alpha_{2}},\dots,\nu_{\alpha_{\lambda}})]$ and
$[(\nu_{\alpha_{1}},
\nu'_{\beta_{2}},\dots,\nu'_{\beta_{\lambda'}})]$ only remain to
be pondered. Then again condition $(c)$ ensures that $\alpha_{2} =
\beta_{2}$ and the product is equal 0 for sure until
$\nu_{\alpha_{2}} \ne \nu'_{\alpha_{2}}$. By repeating the
reasoning over and over again we will come to the conclusion that
$M_{[(\nu_{\alpha_{\kappa}})]}M_{[(\nu'_{\beta_{\kappa}})]} \ne 0$
might be expected only if equalities $\alpha_{\kappa} =
\beta_{\kappa}$ and $\nu_{\alpha_{\kappa}} =
\nu'_{\beta_{\kappa}}$ will be true for all $\kappa =
1,2,\dots,\min{\{\lambda, \lambda'\}}$. However, the chosen
classes become the same for $\lambda = \lambda'$. On the other
hand $\lambda \ne \lambda'$ indicates that one of these classes is
not terminal. This contradiction of the way that classes
$[(\nu_{\alpha_{\kappa}})_{\kappa = 1,2,\dots,\lambda}]$ and
$[(\nu'_{\beta_{\kappa}})_{\kappa = 1,2,\dots,\lambda'}]$ were
defined implies that for every classes belonging to $\mathcal{T}$
an equality
$M_{[(\nu_{\alpha_{\kappa}})]}M_{[(\nu'_{\beta_{\kappa}})]} = 0$
is true. Because any $(q_{j_{\kappa}}) \in [(\nu_{j_{\kappa}})]$
is characterized as the sequence of operation that outcomes
$(\nu_{j_{\kappa}})$ has occurred it follows that:
\begin{equation}
\mbox{Tr}(M_{[(\nu'_{\beta_{\kappa}})]}\rho_{(q_{j_{\kappa}})}) = \left\{\begin{array}{lll}1, & \mbox{if} \quad [(\nu'_{\beta_{\kappa}})] = [(\nu_{j_{\kappa}})];\\
0, & \mbox{if} \qquad ~~~~\mbox{else}.\\
\end{array}\right. \label{trwzor}\end{equation}
By replacing $M_{[(\nu'_{\beta_{\kappa}})]}$ with the payoff
operator $X_{e}$ in (\ref{trwzor}) it leads to convenient
representation of $\widetilde{u}$ as:
\begin{equation}
\widetilde{u} \colon \bigcup(\mathcal{T}) \to \mathbb{R}^n, \quad
\forall_{(q_{j_{\kappa}}) \in [(\nu_{j_{\kappa}})]}
~\widetilde{u}(q_{j_{\kappa}}) = \Delta_{[(\nu_{j_{\kappa}})]}.
  \label{funkcjau}
\end{equation}
\subsection{Quantum extensive game}
The results of previous subsection allow to note:
\begin{definition} A quantum extensive game form $(N, |\Psi\rangle,
\{\mathcal{U}_j\}, \mathcal{H}, \widetilde{P})$ together with
determined utility function (\ref{funkcjau}) is called a quantum
extensive game. \label{quantumextensivegame}
\end{definition}
Now, when we have full description of extensive game played on
qudits we remark upon a role of measurement in these games. In
games played classically it is obvious that an event that we have
just observed after an action taken by a player will be agree with
that action. It also will happen if we put into our scheme an
initial state of the form
$e^{i\varphi}|\nu_{1},\nu_{2},\dots,\nu_{m}\rangle$ simultaneously
with the set of unitary operators for all the players defined by
the equation (\ref{operatorybazowe}). These components define a
game that after a move of each player there is only one
measurement outcome that could be observed. Then expected utility
values for the all players would be one of the vectors
$\Delta_{[\cdot]}$. Notice more that the game is in essence an
extensive game by Definition \ref{egame}. However, in general case
of the quantum game there are more than one outcome that could be
measured on qudit after an action of each player. This feature
formulates the expected utility value as a convex combination of
vectors $\Delta_{[\cdot]}$ in which coefficients are $\Pr(\cdot)$:
\begin{equation}
\widetilde{u}(s) = \sum_{(q_{j_{\kappa}})|s \in
\bigcup(\mathcal{T})}\Pr((q_{j_{\kappa}})|s)\widetilde{u}((q_{j_{\kappa}})|s).
\label{pis}
\end{equation}
where $(q_{j_{\kappa}})|s$ denotes a history consistent with
profile $s$.

In the process of defining the scheme there was no need to relate
the quantum extensive game to an extensive game drawn from
Definition \ref{egame} so far. It point out that both of these
notions are independent of each other, at least in respect of
hints needed to playing this game. However, in order to consider
any connections between classic and quantum games it is
necessarily to ascribe the set of all of quantum extensive games
to those which are quantum realization of fixed classic game. Our
idea coincides with the conception based on papers \cite{eisert}
and \cite{marinatto} which involves identification of each outcome
of a classical static game with exactly one measurement outcome of
quantum system provided for a quantum game. Thus with regard to
our scheme for $(|\Psi\rangle, N, \{\mathcal{U}_{j}\},
\mathcal{H}, \widetilde{P})$ to become a quantum realization of
$(N, H, P, \{\mathcal{I}_{i}\}_{i \in N}, u)$ it is necessary for
components $\mathcal{H}$ and $H$ to be isomorphic. Furthermore,
player functions and utility functions of these games should be
the same for respective arguments.

To specify the thought for a given $(N, |\Psi\rangle,
\{\mathcal{U}_j\}, \mathcal{H}, \widetilde{P}, \widetilde{u})$ let
us consider an arbitrary set $C(\mathcal{H})$ of class
representatives $(q_{j_{\kappa}})|[(\nu_{j_{\kappa}})]$ of all
classes $[(\nu_{j_{\kappa}})]$ from $\mathcal{H}$ that satisfies
conditions $i)$ to $iii)$ of Definition \ref{qextensiveform}. Then
a tuple $(N, C(\mathcal{H}), \widetilde{P}, \widetilde{u} )$ is an
extensive game with respect to Definition \ref{egame}. Moreover,
each of these games does not depend on the choice of a set
$C(\mathcal{H})$. In fact, for any sets $C(\mathcal{H})$,
$C'(\mathcal{H})$ of class representatives, games: $( N,
C(\mathcal{H}), \widetilde{P}, \widetilde{u} )$ and $( N,
C'(\mathcal{H}), \widetilde{P}, \widetilde{u} )$ are isomorphic
via an arbitrary one-to-one mapping $\xi\colon C(\mathcal{H}) \to
C'(\mathcal{H})$ with the property that the image of any class
representative $(q_{j_{\kappa}})|[(\nu_{j_{\kappa}})]$ is a class
representative of the same class $[(\nu_{j_{\kappa}})]$. Taking an
advantage of the above analysis, notion of quantum realization of
game can be defined as follows:
\begin{definition}
Let an extensive game $\Gamma = (N, H, P, \{\mathcal{I}_{i}\}_{i
\in N}, u)$ be given. Every game $(N, |\Psi\rangle,
\{\mathcal{U}_j\}, \mathcal{H}, \widetilde{P}, \widetilde{u})$
such that $(N, C(\mathcal{H}), \widetilde{P}, \widetilde{u} )$ and
~$\Gamma$ are isomorphic is called a~quantum realization of\,
$\Gamma$. \label{quantumrealization}
\end{definition}
From the above definition we can observe that (taking both the
shape of a strategic and an extensive one) has  infinitely many
quantum realizations differing from each other by a given initial
state and (or) unitary operations available to players. On the
other hand the quantum game is the realization of exactly one
classic game (to an accuracy of isomorphism). Notice that the
well-known quantum realization of the Prisoner Dilemma
(\cite{eisert2}, \cite{nawaz}) or the Battle of the Sexes
(\cite{fracor}, \cite{marinatto}, \cite{nawaz-1}) are the examples
of the notion of quantum realization that has been expressed by
Definition \ref{quantumrealization}.
\section{Examples of quantum extensive games}
\begin{figure}[t]
\centering
\includegraphics[angle=0, scale=0.65]{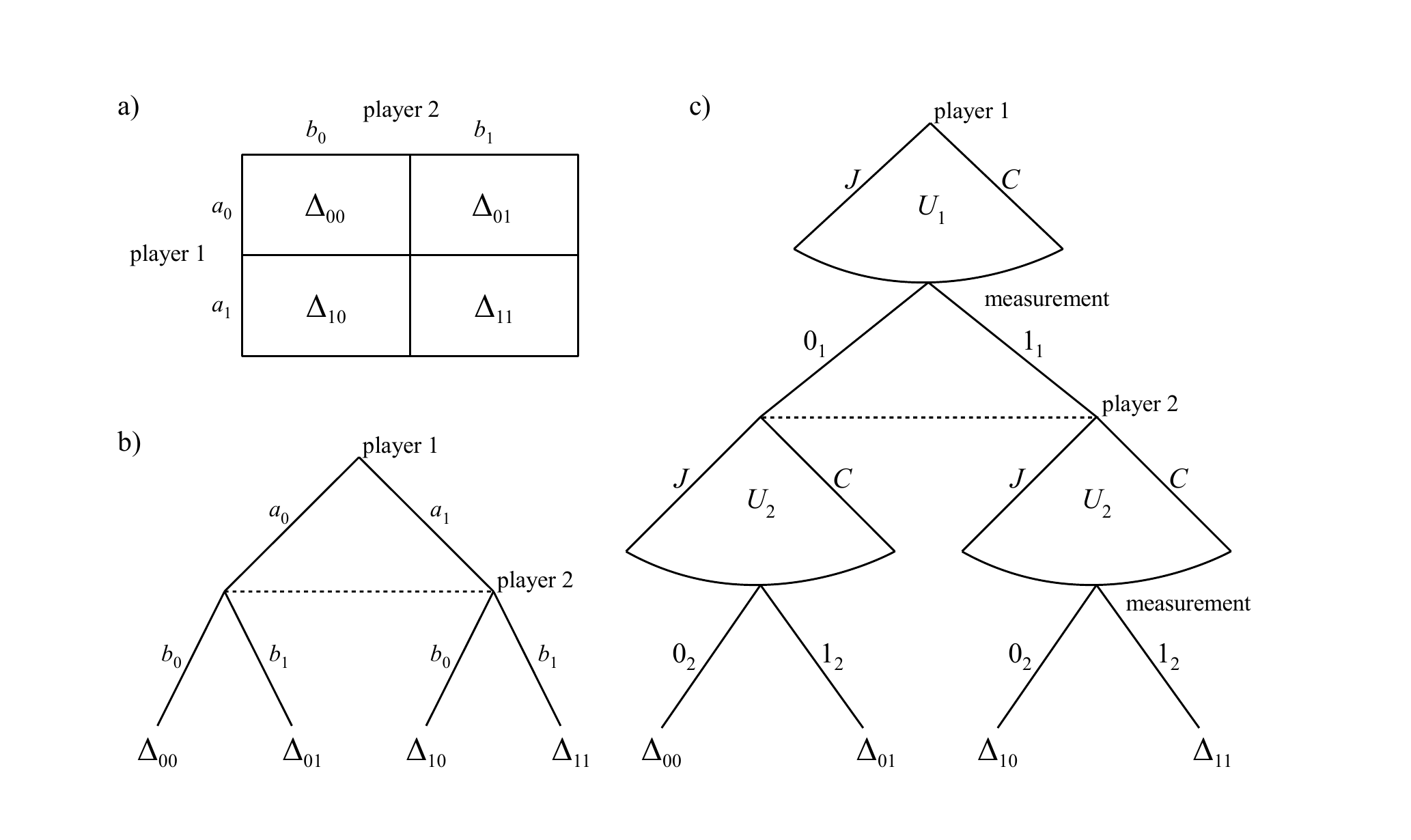}
\caption{Static game (a), its equivalent in the extensive form (b)
and the quantum realization of these game (c). In the quantum
realization each of players chooses among continuum of actions
specifying $\phi$ i $\theta$. The range of possibilities
corresponds to two edges $J$ i $C$ connected with an arc.}
\label{figure1}
\end{figure}

First, we will learn that our scheme of playing extensive games
agree with the scheme described in the second section. A
possibility of such test results from the fact that every static
game (a game in which players make their moves at the same time)
can be depicted in an extensive form. In particular if a
two-player static game in which each of the players has a
two-element action set is given respectively: player 1 has $A =
\{a_{0}, a_{1}\}$, player 2 has $B = \{b_{0}, b_{1}\}$ then one of
the extensive form of this game is following:
\begin{equation}
\Gamma_{1} = ( \{1,2\}, H, P, (I_{i})_{i \in \{1,2\}}, u)
\label{example1}
\end{equation}
{\em in which the components are defined as follows$\colon$}
\begin{enumerate}
\item[{\em(i)}] $H = \bigl\{\emptyset, (a_{0}), (a_{1}), (a_{0},
b_{0}), (a_{0}, b_{1}), (a_{1}, b_{0}), (a_{1}, b_{1}) \bigr\}$,
\item[{\em(ii)}] $P\colon \{\emptyset, (a_0), (a_1)\} \to
\{1,2\}$,~ $P(\emptyset) = 1,~ P(a_{0}) = P(a_{1}) = 2$,
\item[{\em(iii)}] $I_{1} = \{\emptyset\},~ I_{2} = \{(a_{0}),
(a_{1})\}$, \item[{\em(iv)}] $u\colon \{(a_k, b_l)\}_{k,l \in
\{0,1\}} \to \mathbb{R}^2$,~ $u(a_k, b_l) = \Delta_{kl}$.
\end{enumerate}
The static game and its equivalent in extensive form are presented
in the left-hand side of Figure \ref{figure1}. In the extensive
game player 1 moves first (or vice versa). After that according to
the player function player 2 acts. Since information set $I_{2}$
of player 2 (a~dotted line) encapsulates both of the previous
player's actions, this player knows only that it is his turn to
make move na without information which of the two possible
histories has occurred in fact. This situation can therefore be
treated as when both the players carry out their actions at the
same time, as it happens in static games. Assuming that the
respective outcomes of both the games correspond to the same
payoff values, players' decisions about their actions are
indifferent to what of the game they actually play. So, if a
static game and its extensive version are given and a scheme of
quantum static games is correct, the players should be indifferent
to whether they are playing the quantum static game or the quantum
extensive version. The example below shows that the mentioned
property embedded into our scheme.
\begin{example} Let us consider the following quantum extensive
game$\colon$ \label{eexample1}
\begin{equation}
\mathrm{Q}\Gamma_{1} = ( \{1,2\}, |\Psi\rangle, \{\mathcal{U}_1,
\mathcal{U}_2\} , \mathcal{H}, \widetilde{P}, \widetilde{u})
\label{qexample1}
\end{equation}
defined by the components$\colon$
\begin{enumerate}
\item $|\Psi\rangle$ is in the form of (\ref{initialstate}), \item
$\mathcal{U}_{j}$ is a set of unitary operations defined by
(\ref{operator}) i (\ref{operator2}), \item $\mathcal{H} =
\{\emptyset, [(0_1)], [(1_1)], [(0_1,0_2)], [(0_1,1_2)],
[(1_1,0_2)], [(1_1,1_2)]\},$ \item $\widetilde{P}\colon
\{\emptyset, [(0_1)], [(1_1)]\} \to \{1,2\}$,~
$\widetilde{P}(\emptyset) = 1$,
 $\widetilde{P}([(0_1)])=
\widetilde{P}([(1_1)])= \{2\}$,\ \item $\widetilde{u}\colon
\bigcup\left(\{[(\nu_1,\nu_2)]\colon (\nu_{1}, \nu_{2}) \in \{0,
1\}^2\}\right) \to \mathbb{R}^2,$~
$\widetilde{u}\left(\bigcup([(\nu_1,\nu_2)])\right) =
\{\Delta_{\nu_{1}\nu_{2}}\}.$
\end{enumerate}
\end{example}
\noindent The game $\mathrm{Q}\Gamma_1$ is depicted in the
right-hand side of Figure \ref{figure1}. Game $(N, C(\mathcal{H}),
\widetilde{P}, \widetilde{u})$ generated by $\mathrm{Q}\Gamma_{1}$
and game (\ref{example1}) are isomorphic via a bijective map
$\xi\colon H \to C(\mathcal{H})$ presented as:
\begin{equation}
 \xi(h) = \left\{\begin{array}{lll} \emptyset, & \mbox{if} & h = \emptyset;\\
q_{1}|[(\nu_1)], &
 \mbox{if} &
h = (a_{\nu_{1}});\\
(q_1,q_2)|[(\nu_1,\nu_2)],  & \mbox{if} & h = (a_{\nu_{1}},
b_{\nu_{1}}). \end{array}\right.
\end{equation}
According to Definition \ref{quantumrealization} game
$\mathrm{Q}\Gamma_1$ is a quantum realization of game $\Gamma_1$.
Let find out course of the game for an arbitrary strategy profile
$s = (s_{1}, s_{2})$ and then an~outcome corresponds to $s$.

\begin{figure}[t]
\centering
\includegraphics[angle=0, scale=0.65]{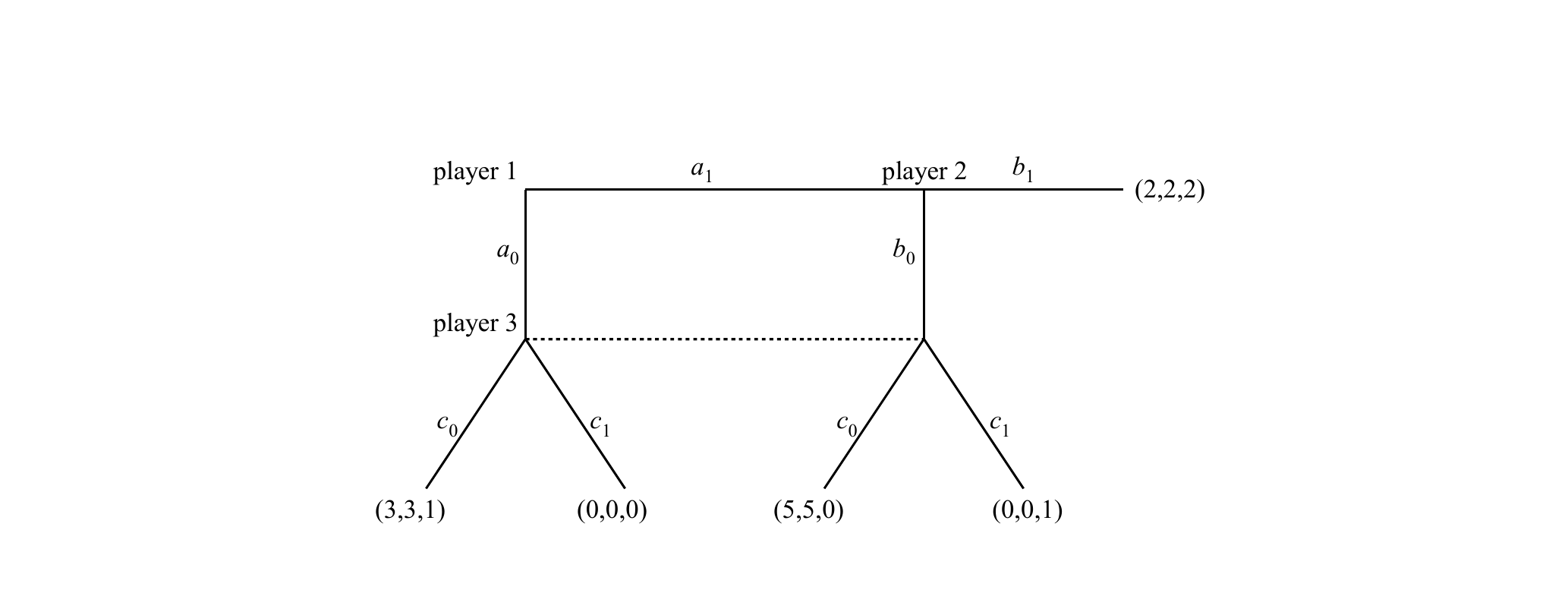}
\caption{The modified Selten's Horse game defined by $\Gamma_{2}$}
\label{figure2}
\end{figure}

Each of the players have only one information set, respectively,
$\mathcal{J}_{1} = \{\emptyset\}$ and $\mathcal{J}_{2} =
\{[(0_1)], [(1_1)]\}$. It implies that profile $(s_{1}, s_{2})$ is
tantamount with action profile $(U_1, U_2)$ where $U_j \in
\mathcal{U}_j$. As the components of $\mathrm{Q}\Gamma_{1}$
dictate, the game starts with unitary operation $U_{1}$ that
player 1 acts on a first qubit of initial state $|\Psi\rangle$:
\begin{eqnarray}
|\Psi_{U_{1}}\rangle &=&
\cos\frac{\gamma}{2}\Biggl(e^{i\phi_{1}}\cos\frac{\theta_{1}}{2}|0\rangle
- \sin\frac{\theta_{1}}{2}|1\rangle\Biggr) \otimes |0\rangle \nonumber\\
&+& \sin\frac{\gamma}{2}\Biggl(\sin\frac{\theta_{1}}{2}|0\rangle +
e^{-i\phi_{1}}\cos\frac{\theta_{1}}{2}|1\rangle\Biggr) \otimes
|1\rangle. \label{obliczenie1}
\end{eqnarray}
Then a measurement on this qubit is preparing. If $M_{0_1} =
|0\rangle \langle0| \otimes I$ then the probability of obtaining
on the first qubit the outcome $0_{1}$ and the state
$|\Psi_{U_{1}}\rangle$ after the measurement are:
\begin{eqnarray}
\Pr(0_1) &=& \cos^2\frac{\gamma}{2}\cos^2\frac{\theta_{1}}{2} +
\sin^2\frac{\gamma}{2}\sin^2\frac{\theta_{1}}{2}, \nonumber \\
|\Psi_{\{U_{1}, 0_{1}\}}\rangle &=&
\frac{1}{\sqrt{\Pr(0_{1})}}\Biggl(
e^{i\phi_{1}}\cos\frac{\gamma}{2}\cos\frac{\theta_{1}}{2}|00\rangle
+ i\sin\frac{\gamma}{2}\sin\frac{\theta_{1}}{2}|01\rangle\Biggr).
\end{eqnarray}
An analogous calculation for $M_{1_1} = |1\rangle \langle1|
\otimes I$ give:
\begin{eqnarray}
\Pr(1_1) &=& \cos^2\frac{\gamma}{2}\sin^2\frac{\theta_{1}}{2} +
\sin^2\frac{\gamma}{2}\cos^2\frac{\theta_{1}}{2}, \nonumber \\
|\Psi_{\{U_{1}, 1_{1}\}}\rangle &=&
\frac{1}{\sqrt{\Pr(1_{1})}}\Biggl(
-\cos\frac{\gamma}{2}\sin\frac{\theta_{1}}{2}|10\rangle +
ie^{-i\phi_{1}}\sin\frac{\gamma}{2}\cos\frac{\theta_{1}}{2}|11\rangle\Biggr).
\end{eqnarray}
After each of histories $\{U_{1}, 0_{1} \}$ and $\{U_{1}, 1_{1}
\}$ it is the player 2 turn now. All histories after which the
second player make move, belong to her information set. It follows
that, she performs an operation $U_{2}$ on the second qubit
regardless of a measurement outcome on the first qubit. If a
couple $\{U_{1}, 0_{1}\}$ occurred, the state becomes:
\begin{eqnarray}
|\Psi_{\{U_{1}, 0_{1}\}, U_{2}}\rangle &=&
\frac{1}{\sqrt{\Pr(0_{1})}}\Biggl(e^{i\phi_{1}}\cos\frac{\gamma}{2}\cos\frac{\theta_{1}}{2}|0\rangle
\otimes \biggl(e^{i\phi_{2}}\cos\frac{\theta_{2}}{2}|0\rangle -
\sin\frac{\theta_{2}}{2}|1\rangle\biggr) \nonumber\\
&+& i\sin\frac{\gamma}{2}\sin\frac{\theta_{1}}{2}|0\rangle \otimes
\biggl(\sin\frac{\theta_{2}}{2}|0\rangle +
e^{-i\phi_{2}}\cos\frac{\theta_{2}}{2}|1\rangle\biggr)\Biggr).
\label{wyrazenie28}
\end{eqnarray}
By use of equations (\ref{rozpiska2}) expression
(\ref{wyrazenie28}) can be rewritten in the form:
\begin{equation}
|\Psi_{\{U_{1}, 0_{1}\}, U_{2}}\rangle = \frac{\chi_{00}|00\rangle
+ \chi_{01}|01\rangle}{\sqrt{\Pr(0_{1})}}. \label{stateiu}
\end{equation}
The probability of getting outcome $\nu_{2} \in \{0,1\}$ and the
post-measurement state given that outcome $\nu_{2}$ occurred are
\begin{equation}
\Pr(\nu_{2}) = \frac{|\chi_{0\nu_{2}}|^2}{\Pr(0_1)}, \quad
|\Psi_{\{U_{1}, 0_{1}\}, \{U_{2}, \nu_{2}\}}\rangle =
\frac{\chi_{0\nu_{2}}|0\nu_{2}\rangle}{\sqrt{\Pr(0_1)}}.
\end{equation}
In case if history $\{U_{1}, 1_{1}\}$ has occurred we get
\begin{equation}
\Pr(\nu_{2}) = \frac{|\chi_{1\nu_{2}}|^2}{\Pr(1_1)}, \quad
|\Psi_{\{U_{1}, 1_{1}\}, \{U_{2}, \nu_{2}\}}\rangle =
\frac{\chi_{1\nu_{2}}|1\nu_{2}\rangle}{\sqrt{\Pr(1_1)}}.
\label{obliczenie2}
\end{equation}
Let determine a payoff vector correspond to $U = (U_{1}, U_{2})$.
Notice first that a set of all terminal histories consistent with
$(U_{1}, U_{2})$ is made up of four histories at most. The form of
this set is
\begin{equation}
\{(q_{1}, q_{2})|U\} = \bigl\{(\{U_{1}, \nu_{1}\}, \{U_{2},
\nu_{2}\})\colon (\nu_{1},\nu_{2}) \in \{0,1\}^2\bigr\}.
\label{ZqU}
\end{equation}
The probability distribution $\Pr(\cdot)$ on the set (\ref{ZqU})
are expressed by the formula $\Pr(\{U_{1}, \nu_1\}, \{U_{2},
\nu_2\}) = |\chi_{\nu_1\nu_2}|^2$ through (\ref{obliczenie1}) to
(\ref{obliczenie2}) and the utility function $\widetilde{u}$
defined on the same domain takes values $\widetilde{u}(\{U_{1},
\nu_1\}, \{U_{2}, \nu_2\}) = \Delta_{\nu_1\nu_2}$. Substituting
the last calculations to the formula (\ref{pis}) the expected
payoffs for the players is as follows$\colon$
\begin{equation}
\widetilde{u}(U) = \sum_{(\nu_1, \nu_2) \in
\{0,1\}^2}|\chi_{\nu_1\nu_2}|^2\Delta_{\nu_1\nu_2}.
\end{equation}
The utility payoffs assigned to $(U_{1}, U_{2})$ are the same as
the one in Eisert's et al scheme.
\begin{example}
Let us consider a three player extensive game given in Figure
\ref{figure2}$\colon$
\begin{equation}
\Gamma_{2} = (\{1,2,3\}, H, P, (I_{i})_{i \in \{1,2,3\}},u)
\label{seltenhorse}
\end{equation}
determined by the following components:
\begin{enumerate}
\item $H = \{\emptyset, (a_0), (a_1), (a_0, c_0), (a_0, c_1),
(a_1, b_0), (a_1, b_1), (a_1, b_0, c_0), (a_1, b_0, c_1)\}$, \item
$P(\emptyset) = 1,~ P(a_1) = 2,~ P(a_0) = P(a_1, b_0) = 3$, \item
$I_{1} = \{\emptyset\},~ I_{2} = \{(a_1)\},~ I_{3} = \{(a_0),
(a_1, b_0)\}$, \item $u(a_0, c_0) = (3,3,1)$, $u(a_0, c_1) =
(0,0,0)$, $u(a_1,b_1) = (2,2,2)$, $u(a_1,b_0,c_0) = (5,5,0)$,
$u(a_1,b_0,c_1) = (0,0,1)$.
\end{enumerate}
\end{example}
It is a modified Selten's Horse game towards the payoffs.
Profiles$\colon$ $(a_0, b_1, c_0)$ and $(a_1, b_1, c_1)$ are Nash
equilibria in this game and each of them could be equally likely
as a scenario of the game indeed. The payoff for players 1 i 2
assigned to
 $(a_0,b_1,c_0)$ is more beneficial than the outcome corresponding to $(a_1, b_1, c_1)$ that is desirable for player 3. The
uncertainty of a result of the game follows from the peculiar
strategic position of player 3. She could try to affect decision
the others through her announcement before the game starts that
she is going to put action $c_{1}$. Then, under the preference of
the others players the history $(a_{1}, b_{1})$ should occur given
that the statement of player 3 is credible enough.

 There are
games among the quantum realizations of game $\Gamma_{2}$ that
have unique Nash equilibrium and that profile would be treated as
reasonable profile for players. One of these games is shown below.
\begin{equation}
Q\Gamma_{2} = (\{1,2,3\}, |\Psi \rangle, \{\mathcal{U}_{1},
\mathcal{U}_{2}, \mathcal{U}_{3}\}, \mathcal{H}, \widetilde{P},
\widetilde{u}) \label{qseltenhorse}
\end{equation}
{\em with initial state:}
\begin{equation} |\Psi\rangle =
\cos\frac{\gamma}{2}|000\rangle +
i\sin\frac{\gamma}{2}|111\rangle, \quad \rm{dla} \quad \gamma \in
(0,\pi) \label{GHZ}\end{equation} {\em and the other components:}
\begin{enumerate} \item[{\em(i)}] $\mathcal{U}_{j} = \{V_{0}, V_{1}\}$ {\em are defined by formula
(\ref{operatorybazowe}),} \item[{\em(ii)}] $\mathcal{H} =
\{\emptyset, [(0_1)], [(1_1)], [(0_1, 0_3)], [(0_1, 1_3)], [(1_1,
0_2)], [(1_1, 1_2)], [(1_1, 0_2, 0_3)], [(1_1, 0_2, 1_3)]\}$,
\item[{\em(iii)}] $\widetilde{P}(\emptyset) = 1$,
$\widetilde{P}([(1_1)]) = \{2\}$, $\widetilde{P}([(0_1)]) =
\widetilde{P}([(1_1, 0_2)])=\{3\}$, \item[{\em(iv)}]
$\widetilde{u}([(1_1, 0_2, 0_3)]) = \{(5,5,0)\}$,
$\widetilde{u}([(1_1, 0_2, 1_3)]) = \{(0,0,1)\}$,
$\widetilde{u}([(1_1, 1_2]) = \{(2,2,2)\}$, $\widetilde{u}([(0_1,
0_3)]) = \{(3,3,1)\}$, $\widetilde{u}([(0_1, 1_3)]) =
\{(0,0,0)\}$.
\end{enumerate}
\begin{figure}[t]
\centering
\includegraphics[angle=0, scale=0.65]{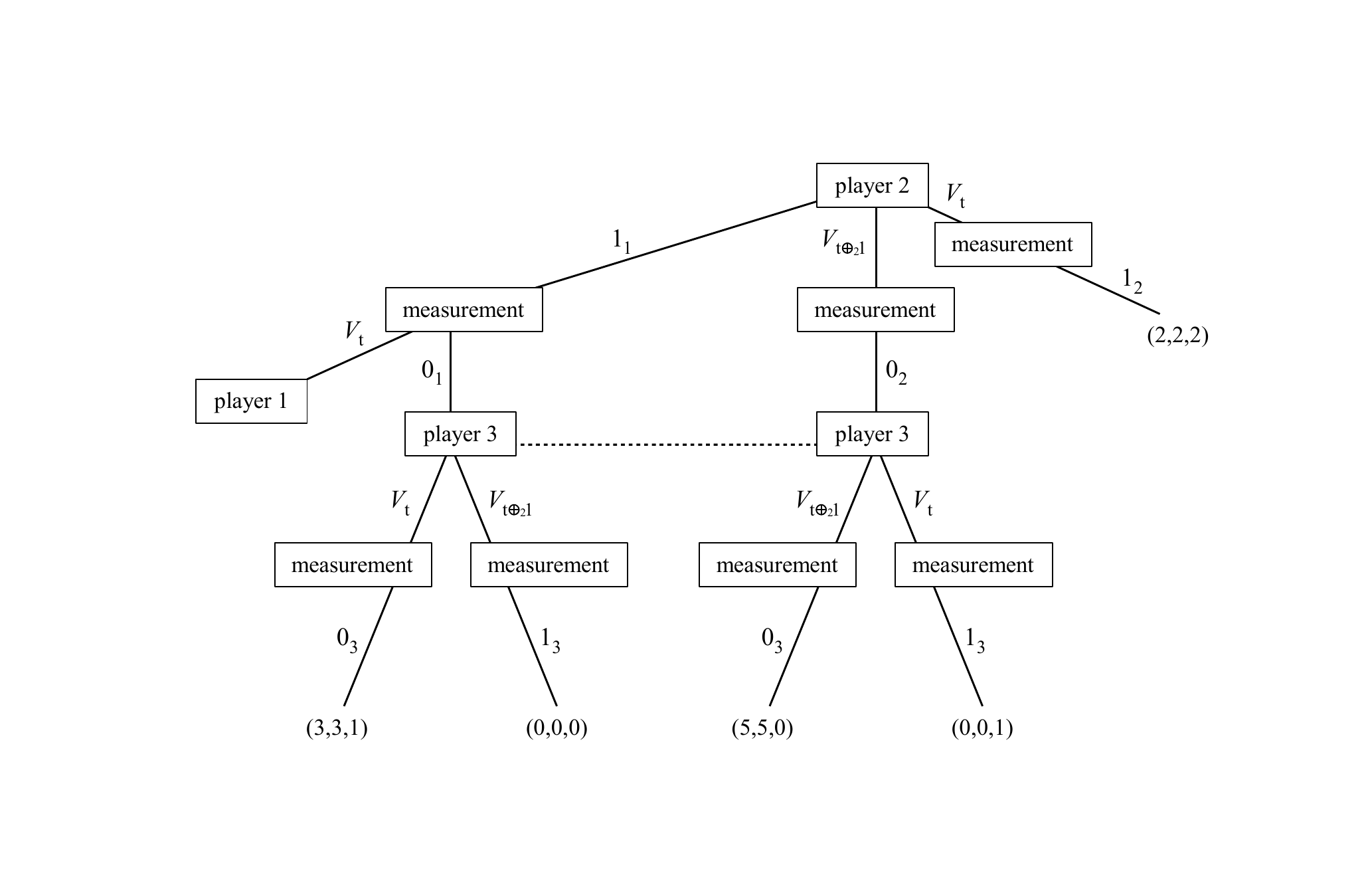}
\caption{Quantum realization of $\mathrm{Q}\Gamma_{2}$. The graph
represents the set of all histories that would occur after action
chosen by player 1} \label{figure3}
\end{figure}
A convenient representation of this game is shown in Figure
\ref{figure3}. Notice first that when player 1 has moved and a
couple $\{V_t, \nu_{1}\}$ has occurred the other players will be
only playing in the `classical' game. It follow form a fact that
after an action of player 1 (she can only apply an identity and
spin-flip operator) and after the measurement the state $\rho$ of
the system collapses to one of the form
$|\nu_{1}\nu_{2}\nu_{3}\rangle \langle\nu_{1}\nu_{2}\nu_{3}|$.
Now, each of the others players acting via $V_t$ on her own qubit
$j$ can get outcome $0_{j}$ or $1_{j}$ with probability equals 1.
Just the same as in games from Example \ref{eexample1} here each
of the players has a~one information set so a set $\{V_{0},
V_{1}\}$ is the set of their strategies. We shall focus on analyze
which of profiles $V$ of these strategies are Nash equilibria. At
first it necessary to  determine expected utility value of each
possible profile of strategies. As an example we will find the
expected payoff for profile $(V_{0}, V_{1}, V_{0})$. This profile
is consistent with two terminal histories: $(\{V_{0}, 0_{1}\},
\{V_{0}, 0_{3}\})$ and $(\{V_{0}, 1_{1}\}, \{V_{1}, 0_{2}\},
\{V_{0}, 1_{3}\})$. To see this, notice that strategy $V_{0}$ of
player 1 is consistent with history $(\{V_{0}, 0_{1}\})$ and
$(\{V_{0}, 1_{1}\})$. If history $(\{V_{0}, 0_{1}\})$ has happened
then $\rho_{(\{V_{0}, 0_{1}\})} = |000\rangle \langle 000|$ and it
is player 3's turn. Her own operation $V_0$ from the profile sets
the history $(\{V_{0}, 0_{1}\}, \{V_{0}, 0_{3}\})$. For this
reason probability $\Pr(\{V_{0}, 0_{1}\}, \{V_{0}, 0_{3}\})=
\Pr(\{V_{0}, 0_{1}\})$ and therefore is equal
$\cos^2\frac{\gamma}{2}$. In a similar way we could confirm that
profile $(V_{0}, V_{1}, V_{0})$ is consistent with terminal
history $(\{V_{0}, 1_{1}\}, \{V_{1}, 0_{2}\}, \{V_{0}, 1_{3}\})$
that occur with probability equal $\sin^2\frac{\gamma}{2}$.
Finally by formula (\ref{pis}) the expected utilities
$\widetilde{u}(V_{0}, V_{1}, V_{0})$ amount to
$(3\cos^2\frac{\gamma}{2},3\cos^2\frac{\gamma}{2},1)$. If all
value $\widetilde{u}(V)$ are at our disposal then by making use of
the inequality (\ref{nashequation}) it turns out that only
profiles of the form $V(t) = (V_{t}, V_{t}, V_{t})$, $t=0,1$ could
be (pure) Nash equilibria. Also, it can be concluded from the set
of equations below:
\begin{equation} \left\{\begin{array}{l}
u_{1}(V(t)) - u_{1}(V_{t \oplus_{2}1}, V_{t}, V_{t}) = -2\cos{(t\pi - \gamma)}, \\
u_{2}(V(t)) -
u_{2}(V_{t}, V_{t \oplus_{2}1}, V_{t}) = 1 - \cos{(t\pi - \gamma)},\\
u_{3}(V(t)) - u_{3}(V_{t}, V_{t}, V_{t \oplus_{2}1}) =
(1+\cos{(t\pi - \gamma)})/2.
\end{array} \right. \label{roznice}\end{equation}
that existence of the Nash equilibria depends on the angle
$\gamma$. This relation can be represented by:
\begin{equation} \mathrm{NE}(\gamma) = \left\{\begin{array}{lll}
V(1), & \mbox{if} & 0 < \gamma \leqslant \pi/2;\\ V(0), &
\mbox{if} & \pi/2 \leqslant \gamma < \pi.
\end{array} \right. \label{roznice}\end{equation}
The expected utilities for the players are
$\widetilde{u}_{1}(V(t)) = \widetilde{u}_{2}(V(t)) = (5 +
\cos{(t\pi - \gamma)})/2$, $\widetilde{u}_{3}(V(t)) = (3 -
\cos{(t\pi - \gamma)})/2$. It can be see now that each player can
gain from playing game $\mathrm{Q}\Gamma_{2}$. Assuming that one
of the equilibria will be chosen in $\Gamma_{2}$, players 1 and 2
can assure oneself 2 utility units and player 3 will get 1 unit
for sure - all are strictly less than utilities from
$\widetilde{u}(V(t))$ irrespective of what a value of $\gamma$
will be. Notice more that there is the unique equilibrium in game
$\mathrm{Q}\Gamma_{2}$ if just $\gamma \ne \pi/2$ or in the case
$\gamma = \pi/2$ the same utilities are assigned to the both
equilibria. It makes the profile $V(t)$ to be considered as
reasonable pair of strategies for players to choose in
$\mathrm{Q}\Gamma_{2}$.
\section{Summary}
Our proposal of quantum playing of extensive game constitutes an
extension of schemes included in papers \cite{eisert} and
\cite{marinatto} as well as their generalizations shown in
\cite{nawaz0} and \cite{nawaz} in the way in which extensive games
broaden static games. Quantum realization of a static game carry
out by means of scheme constructed in third section generates a
game whose set of possible outcomes coincides with the scope of
outcomes that may be  obtained with the use of scheme for
describing quantum static games. It is also a natural assumption
of a game theory to model of static game in its extensive shape in
such a way so that it does not affect the outcome of the game. The
main aim of the research was to defined a new scheme and present
the concept on meaningful examples. Therefore for clarity of
dissertation we restricted to use of concept to only one basic
notion applied to analysis of games - notion of Nash equilibrium.
In reality it is possible to use a tool in form of an arbitrary
equilibrium refinement dedicated to extensive game. Moreover, the
two examples that have been given could be substituted with much
more complicated dynamic games. The concept of quantum extensive
game provides a broad scope of possibilities regarding ways of
analyzing these games and above all way of drawing comparisons
between a classic game and its quantum equivalent.
\section*{Acknowledgments}
The author is very grateful to his supervisor Prof. J. Pykacz from
the Institute of Mathematics, University of Gda\'nsk, Poland for
great help in putting this paper into its final form.

\end{document}